\documentclass[a4paper]{spie}  
\usepackage{psfig}
\usepackage[]{graphicx}

\title{Fresnel lenses for X-ray and Gamma-ray Astronomy } 

\author{Gerry Skinner\supit{a}, Peter von Ballmoos\supit{a},
 Neil Gehrels\supit{b}, and John Krizmanic\supit{b}
\skiplinehalf
\supit{a}
CESR, 9, avenue du Colonel Roche, 31028 Toulouse, France \\
\supit{b}NASA's GSFC,  Code 661, Greenbelt, MD20771, USA }

\authorinfo{Further author information: Send correspondence to G. Skinner\\E-mail: skinner@cesr.fr,
Telephone: 0033 5 61 55 85 61}

\begin{document}

\def \ocross {\odot}
\def \deg {$^\circ$}
\def \muas {$\mu''$}
\def \Msun{M$_\odot$}
\def \Rsun{R$_\odot$}
\def\etal{{\sl et al.}}
\def \Pdot{\.P}
\def \Pdotdot{\"P}

  \maketitle

\begin{abstract}
Phase Fresnel lenses have the same imaging properties as zone
plates, but with the possibility of concentrating all of the
incident power into the primary focus, increasing the maximum
theoretical efficiency from 11\% to close to 100\%. For X-rays,
and in particular for gamma-rays, large, diffraction-limited phase
Fresnel lenses can be made relatively easily. The focal length is
very long - for example up to a million kms. However, the
correspondingly high `plate-scale' of the image means that the
ultra-high (sub-micro-arc-second) angular resolution possible with
a diffraction limited gamma-ray lens a few metres in diameter can
be exploited with detectors having ~mm spatial resolution.

The potential of such systems for ultra-high angular resolution
astronomy, and for attaining the sensitivity improvements
desperately needed for certain other studies, are reviewed and the
advantages and disadvantages vis-\`{a}-vis alternative approaches
are discussed.

We report on reduced-scale `proof-of-principle tests' which are
planned and on mission studies of  the implementation of a Fresnel
telescope on a space mission with lens and detector on two
spacecraft separated by one million km. Such a telescope would be
capable of resolving emission from super-massive black holes on
the scale of their event horizons and would have the sensitivity
necessary to detect gamma-ray  lines from distant supernovae.

We show how diffractive/refractive optics leads to a continuum of
possible system designs between filled aperture lenses and
wideband interferometric arrays.

\end{abstract}


\section{INTRODUCTION}
\label{sect:intro}  

If one can obtain a strong enough signal and if at the same time
diffraction-limited performance can be achieved, then the
high-energy X-ray or gamma-ray parts of the spectrum are the
natural wavebands in which to seek ultra-high angular resolution
(Table \ref{aperture_tab}).

It is remarkable that the same technique seems to be capable of
turning both of these `ifs' into reality. It is even more
remarkable that the optical system needed does not require any
high technology but could be constructed today. The technique
which seems to offer this remarkable combination is that of Phase
Fresnel lenses, a variation on Fresnel zone plates.


\emph{ Fresnel Zone Plates} (FZPs, Figure~1a), as invented by
Soret\cite{soret} , focus radiation by blocking it! More precisely
they block with opaque annuli the Fresnel zones from which the
radiation which would arrive at the desired focus with the wrong
phase. Not surprisingly, they are not very efficient. In addition
to the 50\% of incident radiation which is absorbed, 25\% passes
undiverted and nearly 15\% goes not into the primary focus ($n=1$
for a converging lens) but into secondary focii corresponding to
$n=-1,\pm 3, \pm 5, ...$. The maximum theoretical efficiency  is
$1/\pi^2$, or about 10\%.

\begin{table}
  \caption{Diameter needed to achieve an angular resolution of 0.5 \muas}
  \label{aperture_tab}
    \centering
\begin{tabular}{llll}
\\
\hline \\
 Band  &  Energy    &  Wavelength         & Diameter  \\
 \hline \\
 Radio    &               &     5 mm           &   2.5$\times10^6$ km\\
 Optical  &   2.5 eV     &     500 nm          &   250 km         \\
 X-ray    &   6 keV       &     0.2 nm         &   105 m    \\
 Hard X-ray&  100 keV     &     12 pm          &     6.25 m       \\
 Gamma-ray&  1 MeV        &     1.2 pm         &     62.5 cm   \\
 \\ \hline
\end{tabular}
\end{table}

Although 10\% is not bad compared with some other techniques, one
can do better. As well as independently proposing the FZP,
Rayleigh pointed out the possibility (first demonstrated by
Wood\cite{wood}) of circumventing the absorption  and
`straight-through' ($n=0$) losses in what can be termed the
\emph{Phase Zone Plate} (PZP, Figure~1b).
   \begin{figure}
    \label{fzp_fig}
    \psfig{figure=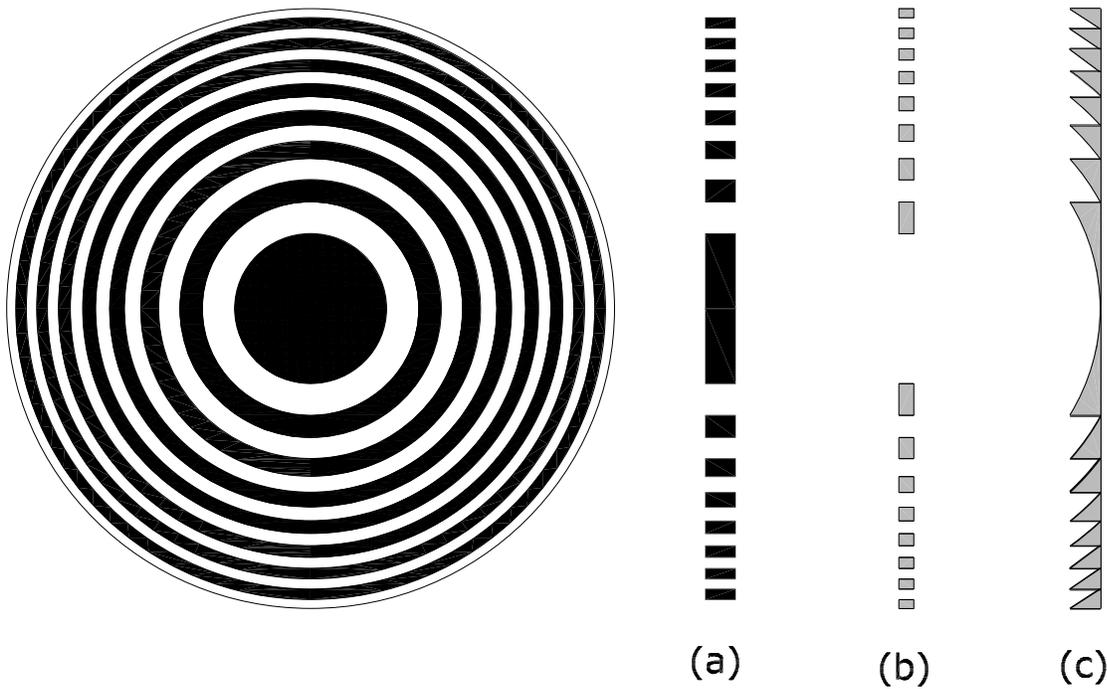,width=16.0cm,angle=0}
    \caption{(a) A Fresnel Zone Plate (FZP), (b) A Phase Zone Plate (PZP)
(c) A Phase Fresnel Lens (PFL)}     \end{figure}
The opaque regions are replaced by ones that allow the radiation
to pass, but which have a thickness of refractive material such as
to impose a phase shift of $\pi$. The theoretical efficiency limit
is raised to $4/\pi^2 =$40.4\%.

Even better, the refractive material can be profiled to give at
each radius in the lens the phase shift such that radiation
arrives at the focal point with \emph{exactly} the correct phase,
not just the nearest multiple of $\pi$. This leads to what we term
here a \emph{`Phase Fresnel Lens'} (PFL, Figure~1c)). This phrase
is adopted here to distinguish such lenses from those originally
invented by Fresnel for use in lighthouses, which are best
considered as simple assemblies of prisms combining light beams
incoherently. Confusingly these incoherent \emph{`Fresnel lenses'}
are distinct from the lenses of which we speak here which rely on
modifying the phase of \emph{`Fresnel Zones'}.

The use of PFLs for X-rays and gamma-rays is possible because all
materials have a refractive index which differs from unity, if
only by a tiny amount. For X-rays and gamma-rays the refractive
index is $\mu=(1-\delta)$, where $\delta$ is small and positive,
corresponding to a refractive index slightly less than one (it
seems to have been Einstein who first pointed this out in
1918\cite{einstein1918}).

The efficiency of a PFL  is limited only by any inaccuracies in
the profile, by absorption or incoherent scattering in the lens
material and by reflective losses at the interfaces. It is easy to
show that if the $rms$ deviations from the ideal profile
correspond to $\lambda/20$ or less, then the resulting losses will
not be greater than 9.9\%. As an example of absorption  in the
lens, consider a PFL made of aluminium working in the first order
at 500 keV. The mean lens thickness will be 0.57 mm, and the
losses in the material only 1.3\%. For energies and materials for
which Compton scattering dominates, the losses are independent of
material and increase in proportion to photon energy. For low $Z$
materials, they are important only below a few keV or above 5--10
MeV. Because the refractive index are always so close to unity,
reflective losses at interfaces are always negligible.

We will discuss here the application of Phase Fresnel Lenses to
astronomy and also consider how certain variations on this
technique allow a continuum of possibilities between a simple lens
and an interferometric array.

\section{ASTRONOMICAL APPLICATIONS}

The astronomical use of X-ray and gamma-ray PFLs has been
discussed by Skinner \cite{iau,paper1,paper2} (see\textsc{} also
the suggestions of Gorenstein \cite{gorenstein} for X-ray
applications).

The possible applications fall into two classes. In the first the
main interest is in the superb angular resolution possible with
this technique. The second class of applications uses PFLs to
overcome the present impasse in gamma-ray astronomy in which
scaling up of existing technologies to improve their sensitivity
is impracticable and maybe even be counterproductive (because
larger systems require larger shields that both lead to worse
dead-time losses and create more neutrons and other secondary
particles).

\subsection{Angular Resolution}

 Among the astronomical objectives which could be achieved using
 the superb angular resolution possible with a high energy PFL
 are:
\begin{enumerate}
  \item Studies of the environment of super-massive black holes in
  the nucleii of active galaxies, on the scale of the event
  horizon.
  \item Investigation of the ejection mechanism in galactic micro
  quasars.
  \item Imaging of active regions on the surfaces of (relatively)
  nearby stars and of the interacting wind systems in close
  binaries.
  \item Visualisation of the explosion of supernovae and of the
  escape of their radioactive by-products.
\end{enumerate}

\subsection{Sensitivity}
    Very high sensitivity is important for the above
    applications but the possibility of having tens of square metres
    of effective area focussed onto a small, low background, detector
    opens up new possibilities even in cases where the objects are
    not resolvable.

    Thus although the radioactive ejecta of a Type Ia supernova could be
    imaged``only"  for those closer than about 50 Mpc ($\sim$ 1 per year),
    the sensitivity of a PFL telescope would make possible the observation  and
spectral study of  many more supernovae. For example gamma-ray
emission should be detectable from Type Ia supernovae out to z=0.1
and the narrower, but much weaker, lines expected from SN II
should be observable to 70 Mpc.

\section{CHALLENGES}

\subsection{Bandwidth}
\label{ss_bandwidth}

Because of the low aperture $d/f$ of practical X-ray and gamma-ray
PFLs, most aberrations are negligible. However, PFLs  are
inherently chromatic, with a focal length which is proportional to
photon energy. The bandwidth beyond which chromatic aberration
will start to degrade the diffraction limited performance is given
by
\begin{equation}
  { \Delta E \over E}  ~\sim ~ 2{\lambda f \over d^2}
~~  ~~ ~~\textrm{or}~~ \Delta E  \sim 3
        \Bigl({{f}\over{10^6~km}}
        \Bigr) \Bigl({{1~m}\over{d}} \Bigr)^2
       ~ keV
         .
\end{equation}
Achromatic combinations of diffractive and refractive lenses have
been suggested\cite{paper2,gorenstein,wang}
but the useful bandpass is still limited.

For narrow line radiation this is not too much of a disadvantage
as the lens may be designed for the relevant energy. Although
ideally the lens thickness should correspond exactly to the
$t_{2\pi}$ for a given energy, in practice a lens designed for an
energy $E$ can be used  over a band (say) 0.7$E$ to 1.5$E$. Thus
multiple lines and the regions around and between them can be
explored with the same lens by adjusting the focal length.

For continuum sources, because of the way the flux is concentrated
onto a small, low background, detector,  one tends to find that
the sensitivity is surprisingly good, despite the fact that only a
small fraction of the incident radiation is used.

Of course radiation at energies either side of the that for which
the system is configured will also arrive at the detector. It will
be less well focussed. For the best angular resolution an energy
resolving detector must be used to ignore such radiation. For
spectral and other studies, though, it can still carry important
information.

Note that whether or not an achromatic combination is used, the
bandpass increases in direct proportion to $f$.

\subsection{Orbital considerations}
\label{ss_orbital}

 For a two satellite system with constant
orientation in inertial space, one of the satellites is
necessarily in non-Keplerian orbit and its position has to be
maintained by a constant thrust\cite{paper1}. Minimising the
gravity gradient forces dictates an orbit well away from the
earth-moon system. Studies\cite{imdc} have shown the feasibility
of a `drift-away' orbit trailing the earth at 1 a.u. from the sun.
The station-keeping forces needed are well within the capabilities
of ion thrusters already available. Alternatively a situation
close to a Lagrangian point could be considered.

\begin{figure}
   \begin{center}
   \begin{tabular}{c}
  \vspace{-1.0cm}
   \includegraphics[height=8cm,angle=0]{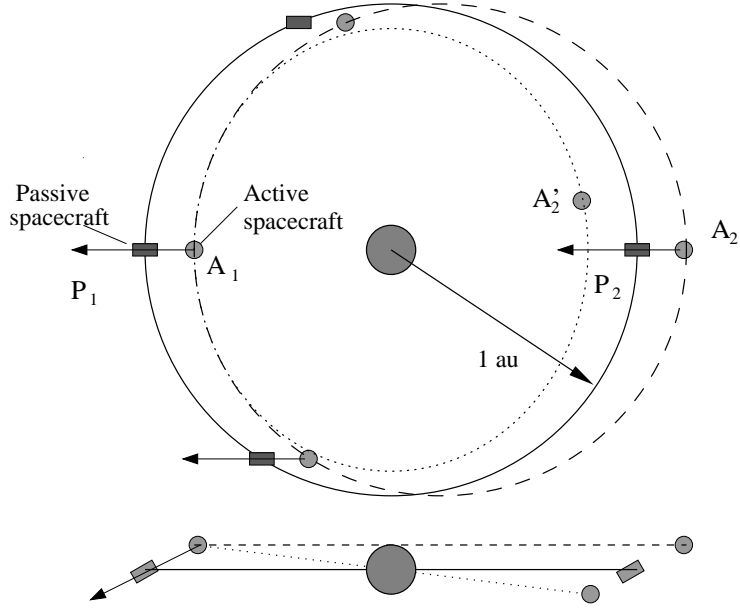}
   \vspace{5mm}

   \end{tabular}
   \end{center}
   \caption[example]
   { \label{orbit-fig}
     To maintain the relative position of two spacecraft in celestial space
     when each is in orbit around the sun requires a thruster on the active spacecraft
    \textsf{ A}.
     In the absence of such a thrust, the spacecraft which
     initially had the correct relative position and velocity at \textsf{ A}
     would arrive not at \textsf{A$_2$} but at \textsf{A$_2'$}
     }
   \end{figure}

The forces necessary to accelerate and decelerate one of the
spacecraft to change the telescope pointing are also an important
consideration. With careful mission planning and allowing a few
weeks for a major repointing, they are comparable with the
station-keeping forces. This is not surprising if one considers
that simply shutting off the station-keeping thrusters would lead
to vector joining the two spacecraft reversing on a time-scale of
6 months (Figure \ref{orbit-fig}).

\subsection{Pointing knowledge}
\label{ss_pointing}

Perhaps the most crucial consideration is the issue of
determination of the pointing direction (as defined, for example,
by the line joining the centre of the lens to the centre of the
detector). The problem the same for any for any ultra-high
resolution telescope -- one has to know where the target is and
where the telescope is pointing, and one has to know it with an
accuracy better than the field of view and with a stability better
than the required resolution.

Even though centroiding allows the determination of positions of
bright stars much better than the resolution of a sensor, Table
\ref{aperture_tab} shows that we are far from the domain of a
simple optical startracker. Studies of "Super-Startrackers" have
already started in the context of
MAXIM\cite{gendreau03,imdc-maxim}.

\section{LONGER IS BETTER?}
 The focal
length of a PFL (or equally a FZP or PZP)  can be expressed as
\begin {equation}
\label{f_eqn}
 f =\Bigl({{p}\over{1~mm}} \Bigr) \Bigl({{d}\over{5~m}}
 \Bigr) \Bigl( {{E}\over{500 ~keV}} \Bigr)\Bigl( {1\over m} \Bigr)  ~1.0\times 10^6~km,
\end {equation}
where $E$ is the operating energy, $p$ is the pitch of the pattern
near the periphery and $d$ is the diameter of the lens. Here we
allow for the fact that higher orders of diffraction, $m$ can in
principle be used.

The diffraction-limited angular resolution is
\begin {equation}
\label{theta_eqn}
 \theta_d~ = ~1.22 {\lambda\over d} ~=~0.12 ~ \Bigl( {{500 keV}\over{E}}\Bigr)~\Bigl({{5~m}\over{d}} \Bigr)
 ~~ micro~arcsec,
\end {equation}

 A natural and frequently asked question is whether there is not some
way of reducing the focal length of a Fresnel telescope system.
Reductions from the extreme values of ~10$^6$ km, which is
suggested by the values used for normalising the parameters in
Equation \ref{f_eqn}, are certainly possible. With some sacrifice
in angular resolution either $d$ or $E$ can be decreased. More
importantly, reducing $p$ will reduce the focal length in
proportion.

The only limits on reducing $p$ are practical ones. The thickness
of the lens, $t$, will be fixed by the working energy and by the
material chosen : $  t = m t_{2\pi} $ where
\begin {equation}
\label{t_eqn} t_{2\pi} = \lambda/(1-\mu) =\lambda/\delta \simeq
\Bigl( {E\over{500~keV}}\Bigr)\Bigl( {{1 ~g~cm^{-3}}\over{\rho}}
\Bigr)   ~ 3.1~mm.
\end {equation}
The approximate form given here for $t_{2\pi}$ in terms of density
$\rho$ is only valid well above any absorption edges. It assumes a
number of electrons per nucleon of 0.48, as for Al, but this ratio
is in the range 0.38--0.5 for all elements Z$>$4. Reducing $p$
leads to structures with higher and higher aspect ratio $t/p$.

The focal length can also be reduced by using $m>1$, but for a PFL
this means that the thickness will be increased and again the
aspect ratio will be correspondingly higher.

Accepting that shorter focal lengths are possible, are they
desirable? From a spacecraft point of view -- yes. The thrust
necessary to overcome gravity gradient forces and maintain a
non-Keplerian orbit is usually proportional to the separation (it
can increase even faster if the system has been positioned near a
Langrangian point to minimise gravity gradient effects). Also the
fuel and/or time needed to reorient the telescope by manoeuvering
one satellite around the other will be higher for large
separations.

The field of view for a given detector size will of course also be
improved for a relatively short focal length.

 On the other hand from the point of view of the performance of the
telescope, there is a lot to be said for long focal lengths :-
\begin{itemize}
  \item {Chromatic abberation:} As discussed in section
  \ref{ss_bandwidth}, the bandpass is proportional to $f$.

  \item {Plate scale:} Particularly at high energies, the spatial
  resolution of detectors is limited. Even when a gamma-ray  photon is
  stopped in a single photo-electric interaction, the resulting energy deposit
  is spread over the path length of the electron in the detector material.
  For a simple PFL (or FZP/PZP)  the spatial
  focal spot size corresponding to diffraction limited angular
  resolution is $f\theta_d$ or  $0.61 p/m $. Thus little is
  gained by reducing $p/m$ beyond a few times the detector
  resolution.
  \item {Navigation:} A possible solution to the problem of
  pointing determination is to measure the  positions of
  the two satellites and so deduce the direction of the vector
  joining them. Spacecraft positions are routinely determined to a
  few meters.  Unless one has a detector some 10 m or more in
  size, this by itself is not sufficient to bring a specific
  object within the field of view. But by using techniques similar
  to those of the GPS, there is a real prospect of improving the
  knowledge of the relative position of the two to the centimeter
  or millimeter scale. With a focal length of $10^6$ km,
  0.5 cm precision would lead to a determination of telescope orientation at
  the 1 \muas\ level. If the focal length is reduced too much
  this approach is no longer viable.

\end{itemize}

\section{GROUND TESTING}

One problem with any space-based optical system, focussing or
interferometric, which has components separated by km or more, let
alone by a million km, is `how can it be tested?'. In fact there
are two related problems - proof of concept and testing of
prototype/engineering-model/flight hardware.

It might be argued that testing of the X-ray/gamma-ray performance
of a phase Fresnel lens system is unnecessary -- that metrology,
together with a knowledge of the index of refraction of the
material(s) used would be sufficient. The metrology is not
especially demanding - for a typical long focal length gamma-ray
Fresnel lens the profiling has a finest scale of the order of
millimeters and  so the $\lambda /20$ optical precision considered
above corresponds to $\sim$ 50 $\mu$m. The index of refraction can
be directly calculated from scattering factors. Detailed
computation of these factors are available \cite{kissel95} and
have been verified against measurements in certain cases
\cite{basavaraju95}. The combinations of interest will,
furthermore, be among those for which the calculations are most
accurate -- scattering in the forward direction by light atoms,
well above absorption edges.

However, we take the view that it would be unwise to invest in a
major space mission without `proof of concept' measurements and we
have been pursuing two general paths for demonstrating the
principles involved.

 \subsection{ Testing miniaturised lenses}

The first approach is to test a scaled down system.  Equation
\ref{f_eqn} shows how the scaling  dependends on the system
parameters and Table \ref{scaling_table} gives some example values
which correspond to testing with the focal length reduced by a
factor of $\sim$2 million. Even this is not enough to bring the
total source/detector distance within the range of vacuum test
facilities, so we propose stacking (say) four lenses in series to
increase their power.  Although the angular resolution will not be
as good as the sub-micro arc-second performance of the final
system it should nevertheless be  about 1 milli arc-second.

\begin{table}
  \caption{Scaling for miniaturising Fresnel lenses for ground testing}
  \label{scaling_table}
    \centering
\begin{tabular}{lccccc}
\\
\hline \\


             &    & Effect on              & \multicolumn{2}{c}{Example value} & Factor      \\
             &    & focal length $f$       & Flight                &   Ground  &             \\
  \hline
  \\
  Energy                & $E$     & $\propto E $     & 500 keV    & 60 keV     &    8.3      \\
  Lens diameter         & $d$     & $\propto d $     &    5 m     & 5 mm       &   1000      \\
  Finest pitch          & $p$     & $\propto p^{-1} $& 1.0 mm     &  4 $\mu$m  &     250      \\
 Total scaling          &         &                  &            &            & $2\times 10^6$\\
 Focal length           &   $f$   &                  & $10^6$ km  &     500 m  &             \\
 Diffraction limit      &$\theta_d$&                 & 0.12 micro arcsec &     1 milli arcsec  &             \\
 Lenses  in series      &  $N$    &                  & 1          & 4          &             \\
 Source-lens distance   &  $u$    &                  &   $\infty$   &      250 m &             \\
 Lens-detector distance &  $v$    &                  &  $10^6$ km  &     250 m &             \\
  \\ \hline
\end{tabular}

\end{table}

 \subsection{ Interferometric testing of gamma-ray lenses}

The most promising way of testing a more representative lens seems
to be by using interferometric techniques.

The action required of a lens is to change the radius of curvature
of the incoming radiation. In the case of focussing a parallel
beam it is changed from infinity to a curvature (positive by
convention) corresponding to converging radiation. Strongly
diverging radiation ($-f<<R<0$) falling on the same lens will
become slightly less diverging.
   \begin{figure}
   \begin{center}
   \begin{tabular}{c}
   \includegraphics[height=9cm]{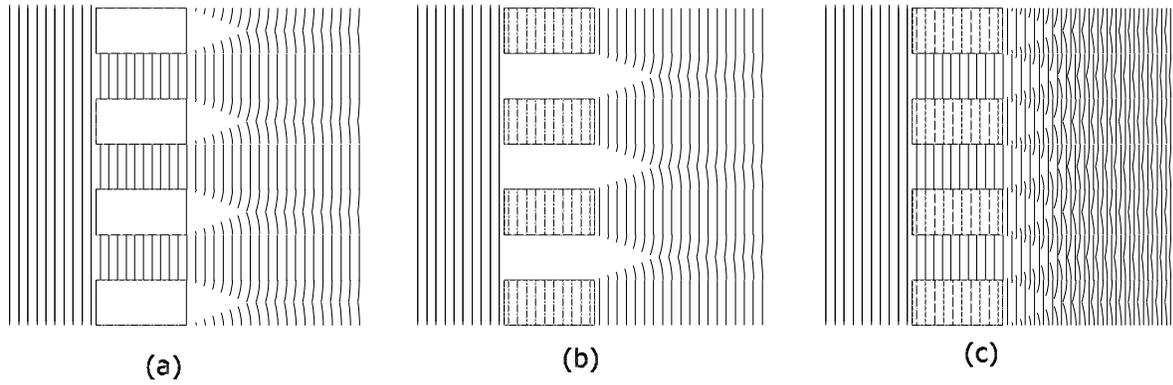}
   \end{tabular}
   \end{center}
   \caption[example]
   { \label{slit_fig}
    Principle of the proposed interferometric testing of gamma-ray
    PFLs.
      (a) Wavefronts falling on a body perforated with slits will form
      again after a sufficient distance on the far side. (b) If
      the material forming the slits is transparent, there will be
      a second component which passes via the material and which
      suffers a phase shift. (c) The net effect (in the forward
      direction) is destructive interference if the phase shift in
      $(2n+1)\pi$.}
   \end{figure}
We propose to detect this change in radius by gamma-ray
interferometry. To create an interference pattern, the  outgoing
radiation must be combined with a sample of the incoming
radiation. This can be done if narrow slits are made in the lens
(Figure \ref{slit_fig}).  Sufficiently far beyond the lens, the
diffractive spread will be such that the radiation passing through
the slits and that passing through the lens will have become
mixed. For slits of width $s$, typical separation $p$, this will
happen for distances greater than about $d_m = s p/\lambda$. Where
the lens has the thickness $t=n t_{2\pi}$, the lens will be like a
transparent window, where $t=(n+1/2) t_{2\pi}$, the transmitted
flux will go to zero. Of course, the photons and their energy do
not simply disappear, they are diverted into orders $m\neq 0$ of
the phase modulating aperture formed by the slits. If the slits
are periodic with $p=2s$, they simply form a diffraction grating
and the energy will go mainly into orders $m=\pm 1$, diverted by
angles $\theta=\pm \lambda /p$.
   \begin{figure}
   \begin{center}
   \begin{tabular}{c}
   \vspace{-3cm}
   \includegraphics[height=9cm]{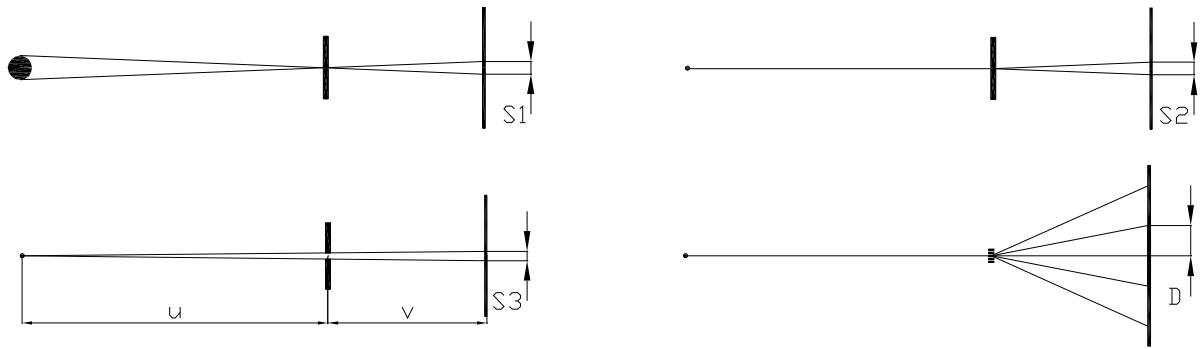}
   \end{tabular}
   \end{center}
   \caption[example]
   { \label{beam_spread_fig}
      The effects which combine to produce the interference pattern
      from a single slit stack of $N$ slits during the proposed interferometric
      testing  of a very long focal length PFL.
      i) The spread, $S_1$  due to a finite source,
      ii) The diffractive spread, $S_2$ following passage through the aperture
      corresponding to the stack of $N$ slits
      iii) The spread, $S_3$ due to geometric spreading
      iv) The splitting into diffractive orders}
   \end{figure}

\begin{figure}
   \vspace{-2.5cm}
   \hspace{8mm}
   \includegraphics[height=7cm, angle=90]{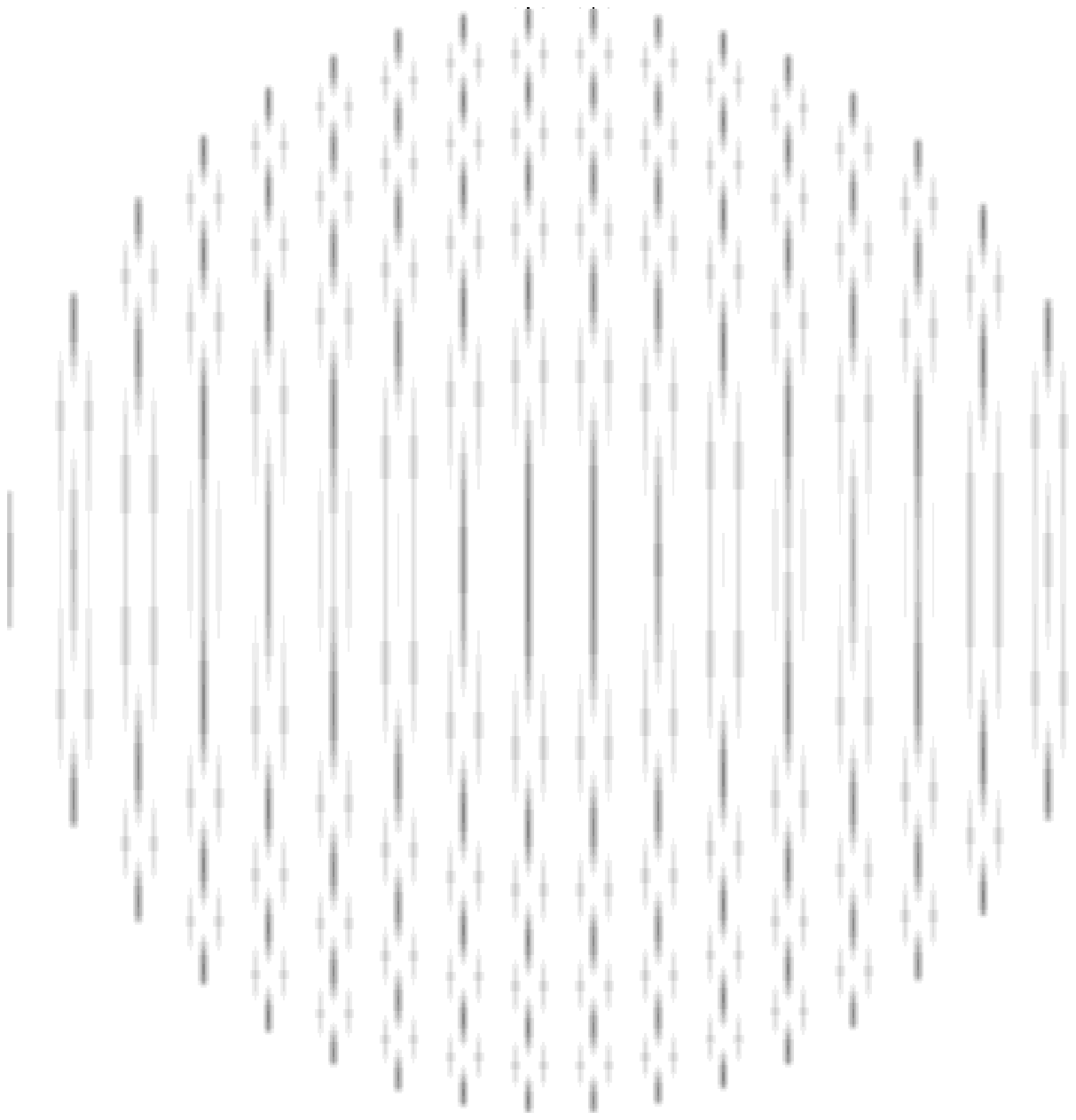}
   \hspace{8mm}
   \includegraphics[height=7cm, angle=90]{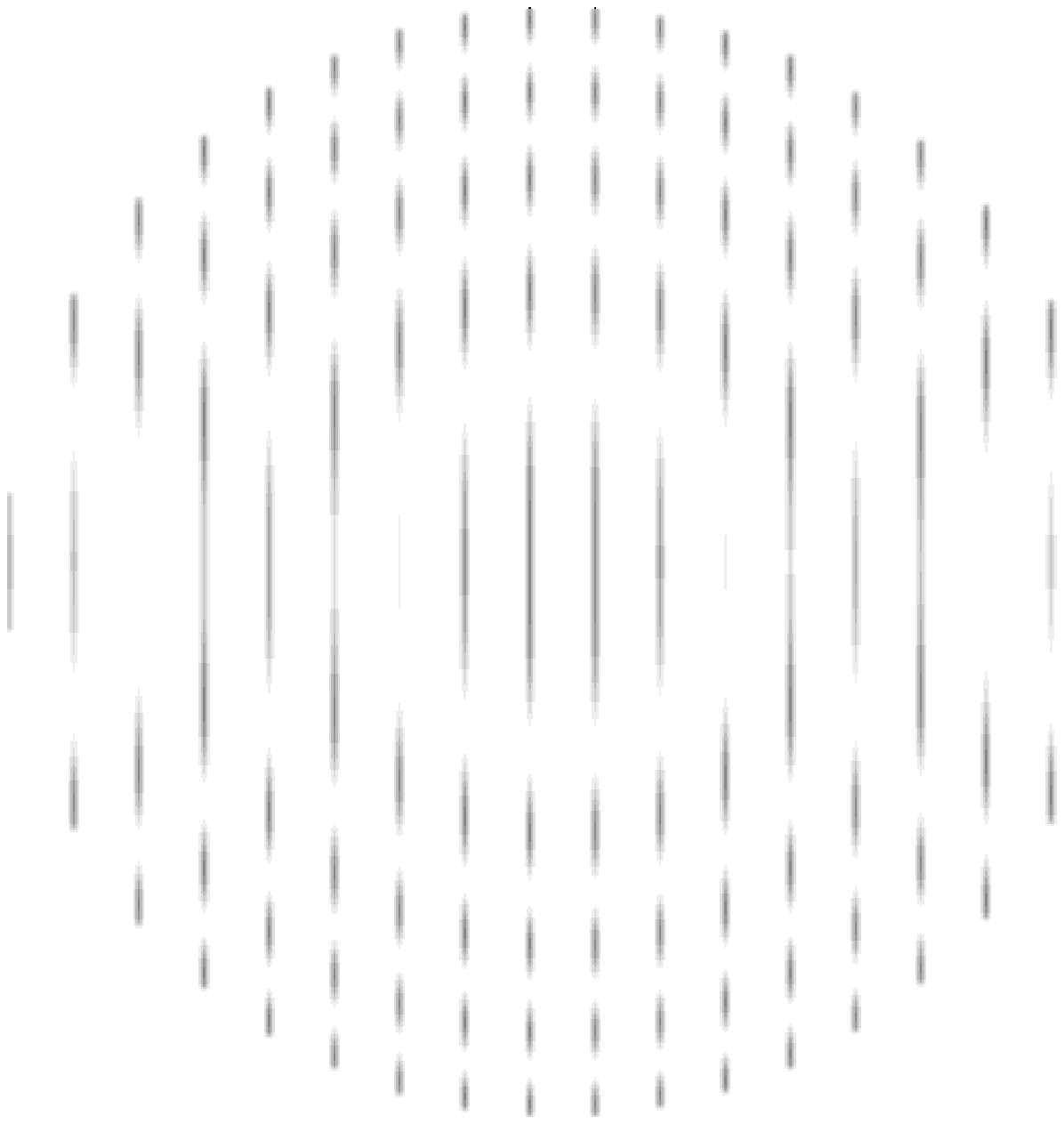}
   \\
   \vspace{-10mm}
     { \hspace{5mm} \includegraphics[width=8.5cm ,trim=0mm 8mm 0mm 8mm]{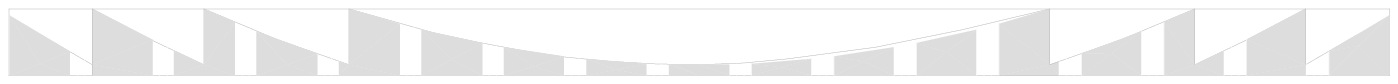}}
   \\
   \vspace{3mm}

   \caption[example]
   { \label{fringepat}
      Fringe patterns from a simulation of an interferometric test of a 500 keV Phase Fresnel
      Lens of focal length $10^5$ km. Test conditions as in Table \ref{fringe_tab}.
      The region shown would be 100 mm in diameter.
      Left:      The pattern with the diffracted beams.
      Right: The same pattern with the diffracted beams blocked or ignored. To make the
      structure more apparent and to simulate the effect of finite detector resolution
      the patterns have been convolved with a 0.6 mm blurring function.
      Below: Cross-section  through the modelled lens.}
   \end{figure}

For this effect to be detectable, one has to observe the pattern
far enough away that the beams are well separated, that is to say
they have been deflected from the straight-through, $m=0$,
direction by more than their width \textit{i.e.}, $Np$, where
there are $N$ slits. Table \ref{fringe_tab} shows that with
possible parameters the deflection can be nearly 10 times $Np$.
Simulations of the fringe patterns expected are shown in Figure
\ref{fringepat}.

Making 0.5 micron wide slits in material several hundred microns
thick would be difficult, so we are proposing to make parts of the
lens from alternating high density/low density materials, with the
low density material taking the place of the slits.

 Ideally the test would be done with a vacuum
beamline, but we are considering the possibility of using a
flexible tube filled with Helium or even accepting the high
absorption and scattering losses associated with a test in open
air. Table \ref{longbeam_tab} illustrates some possibilities.

\begin{table}
  \caption{Parameters for interferometric testing of a gamma-ray Fresnel lens}
  \label{fringe_tab}
    \centering
\begin{tabular}{llll}
\\
\hline \\

  Parameter  &              & \multicolumn{2}{c}{Example value}  \\
  \hline
  \\
  Energy     & $E$          & 500                 & keV\\
  Wavelength & $\lambda$    & $6\times 10^{-12}$  & m \\
  Slit width & $s$          & 0.5                 & $\mu$m \\
  Slit period & $p$         & 1.0                 & $\mu$m \\
  Number of slits & $N$     & 40                   &      \\
  Observing distance & $v$  & 500                  & m    \\
  Source distance & $u$    & 500                 & m    \\
  Assumed source size & $w_s$ & 100                 &  $\mu$m    \\
  Minimum distance for mixing & $d_m=~sp~/\lambda$                        &  0.20         &  m     \\
   Displacement of beam by diffraction &  $D = \lambda v/p$             &       1.24    & mm     \\
  Spread due to finite source size & $S_1=w_s (u+v)/u  $            &        0.10   & m      \\
  Spread due to diffraction        & $S_2= \lambda d /(Np)   $          &      0.03     & mm     \\
  Spread due to width of beam ($N$slits) & $S_3=N p ~(u+v)/u $       &      0.08     & mm     \\
  Total spread                     & $S=(S_1^2+ S_2^2+S_3^2)^{1/2} $    &      0.13     &   mm   \\
  FOM = Displacement/Spread          &$ D/S $              &         9.4   &        \\
  \\ \hline
\end{tabular}

\end{table}

\begin{table}
  \caption{Possible combinations for a `long-beam' interferometric lens test}
  \label{longbeam_tab}
    \centering
\begin{tabular}{clccc}
\\
\hline \\
 Energy &  Source    &  Distance       & Medium  & Transmission  \\
  (keV) &             &      (m)       &         &     (\%)     \\
 \hline \\
 100  &   X-ray generator &   500      &  Vacuum  &  100
 \\
 500   &  Betatron     &     1000       &   He    &   21        \\
 1000 &   Betatron     &      500       &   Air   &   2.2        \\
 \\ \hline
\end{tabular}
\end{table}

\section{A CONTINUUM OF POSSIBILITIES}

The most studied approach to getting ultra high angular resolution
in high energy astronomy is through X-ray interferometry. The
ideas presented here represent an alternative  but the two lines
of approach are not entirely distinct. With multiple mirror
assemblies concentrating flux onto a single detector assembly,
where they all constructively interfere at a single point, the
MAXIM concept is not very different from a focussing optical
system with an unfilled aperture.  Likewise, a PFL could provide a
useful performance with only part of the aperture effective.

\begin{figure}
     \begin{center}
   \includegraphics[height=10cm, angle=0]{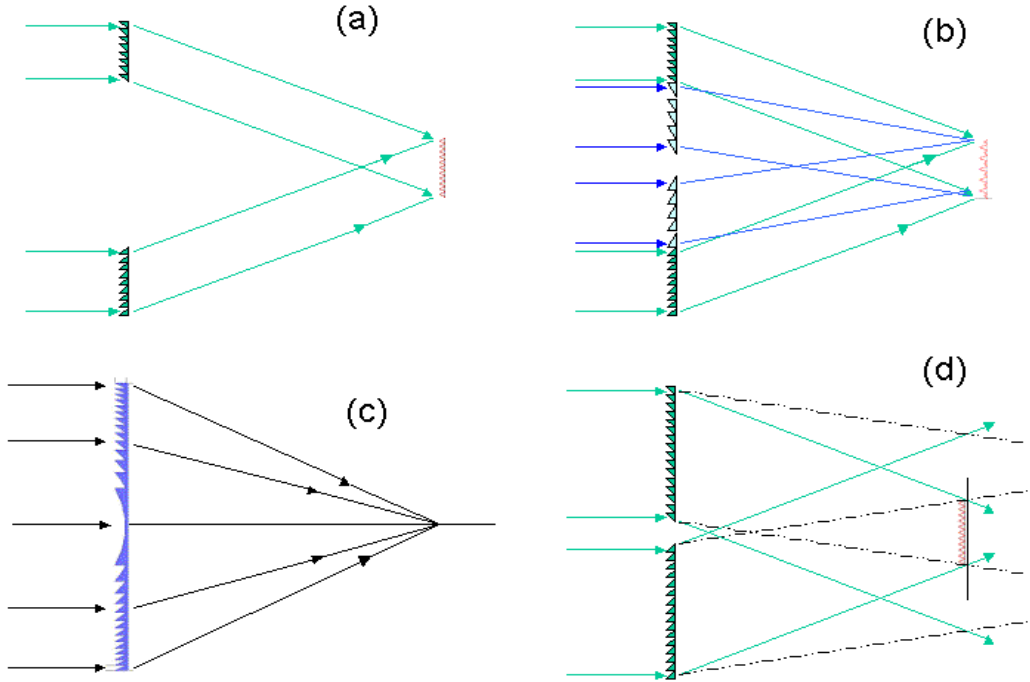}
   \end{center}
   \caption[example]
   { \label{interferom-fig}
      (a) An interferometer with a refractive/diffractive beam
      combiner. (b) Adding a second system with a different pitch.
    (c) Continuous variation of pitch recovers the PFL concept.
    (d) If the pitch is kept constant a very wideband interferometer
    results.}
   \end{figure}

Skinner\cite{paper2} has suggested that a diffractive/refractive
component could be used in a beam combiner in an interferometer
for X-rays or gamma-rays (Figure \ref{interferom-fig}a). In fact,
the region over which flux is collected could be increased by
adding more area as in Figure \ref{interferom-fig}b with a
different pitch to direct the same wavelength towards the
detector. A natural extension would be to have not just two
periods, but a graded pitch -- and one arrives again at a PFL
(Figure \ref{interferom-fig}c).

Alternatively, the pitch could be kept constant, as in a
diffraction grating (Figure \ref{interferom-fig}d).  This last
arrangement is particularly interesting. Of the radiation arriving
with a particular off axis radius $r$, there will always be one
wavelength for which the deflection angle leads to an interference
pattern in the detector plane.  The interference pattern formed is
\textbf{independent of wavelength} -- for two diffractors, it will
just a set of parallel  fringes of the form $Sin^2(2\pi h/p)$,  
where $h$ is the  distance from the axis of a point in the
detector plane. The pitch is simply equal to one half $p$,
independent of wavelength. If the system has axial symmetry, the
radial dependence of the wave amplitude will be a zero order
Bessel function, $J_0(2\pi h /p)$, and so the power will be
$J_0^2(2\pi h /p)$. Again the scale is independent of photon
energy. Basically what is happening is that longer wavelengths are
diffracted from parts of the diffractor at larger radii, and so
the wavefronts arrive at the detector with a larger  angle between
the two interfering wavefronts. The resulting interference pattern
always has the same scale

\begin{figure}
   \includegraphics[width=16cm, angle=0]{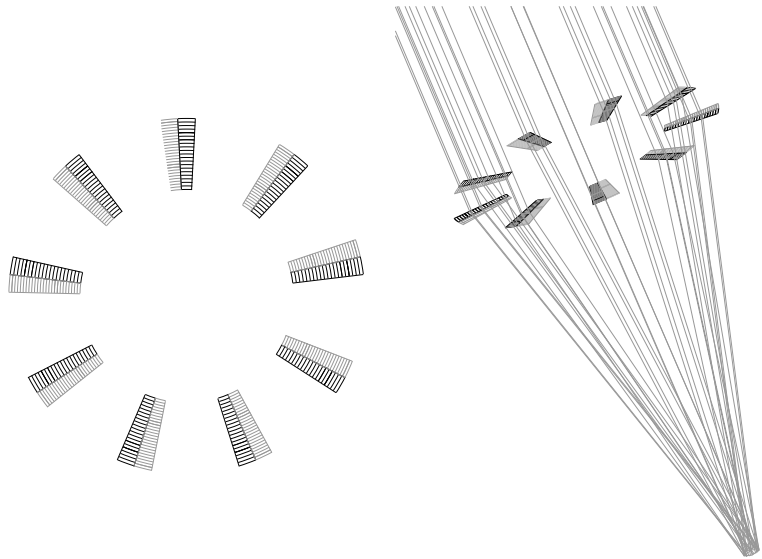}
   \caption[example]
   { \label{minim-fig}
      (a) Use of diffractive beam recombiners in a MAXIM-like
      configuration. The two levels of grey indicate parallel
      systems covering different (perhaps adjacent) bands. (b)
      The converging beams from one set of diffractors at a
      particular energy.
       }
   \end{figure}

 It is easy to imagine a system in which a number of such
diffractors are used in a configuration similar to that of the
MAXIM baseline (Figure \ref{minim-fig}). Assuming that the
diffractors are much larger than the detector plane, the range in
energy will be approximately $E_{max}/E_{min}=d_{max}/2d_{min}$.
So for a 10\% bandpass, they need to be $\sim 0.05d_{max}$ in
length, where $d_{max}$ will be dictated by the aperture size
corresponding to the required resolution and the longest
wavelength. If necessary, the bandpass can be improved by placing
diffractors for adjacent bands alongside the original ones. With a
sufficiently large number of diffractors, the response will
approximate the $J_0^2$ one for a complete annulus.

Although the optical elements would be large, they would be
relatively easy to manufacture,  tolerances and alignment
requirements would be very easily achieved, and there would be no
loss in efficiency associated with  reflections at multiple
surfaces.

Between a filled aperture system with pattern pitch following the
$1/r$ rule for a zone plate scheme as in Figure~\ref{fzp_fig} and
a sparse array with constant pitch diffractors (Figure
\ref{minim-fig}), a continuum of possibilities can be imagined.
Further studies are certainly needed to identify the most
interesting combinations.


\bibliography{aamnem99,spie_2003_gks}   
\bibliographystyle{spiebib}   


\end{document}